\documentclass[aps,pre,preprint,superscriptaddress,showpacs,amssymb,floatfix]{revtex4}
\usepackage{graphicx}
\usepackage{graphics}
\usepackage{amsmath}
\usepackage{tabularx}
\usepackage{color}
\usepackage{doi}
\usepackage{textcomp}
\newcolumntype{Y}{>{\centering\arraybackslash}X}

\begin{document}

\title{Diffusive Majority Vote Model}

\author{J. R. S. Lima}
\affiliation{Departamento de F\'{\i}sica, Universidade Federal do Piau\'{i}, 57072-970, Teresina - PI, Brazil}
\author{F. W. S. Lima}
\affiliation{Departamento de F\'{\i}sica, Universidade Federal do Piau\'{i}, 57072-970, Teresina - PI, Brazil}
\author{T. F. A. Alves}
\affiliation{Departamento de F\'{\i}sica, Universidade Federal do Piau\'{i}, 57072-970, Teresina - PI, Brazil}
\author{G. A. Alves}
\affiliation{Departamento de F\'{i}sica, Universidade Estadual do Piau\'{i}, 64002-150, Teresina - PI, Brazil}
\author{A. Macedo-Filho}
\affiliation{Departamento de F\'{i}sica, Universidade Estadual do Piau\'{i}, 64002-150, Teresina - PI, Brazil}

\date{Received: date / Revised version: date}

\begin{abstract}

We define a stochastic reaction-diffusion process that describes a consensus formation in a non-sedentary population. The process is a diffusive version of the Majority Vote model, where the state update follows two stages: in the first stage, spins are allowed to hop to neighbor nodes with different probabilities for the respective spin orientation, and in the second stage, the spins in the same node can change its values according to the majority vote update rule. The model presents a consensus formation phase when concentration is greater than a threshold value, and a paramagnetic phase on the converse for equal diffusion probabilities, i.e., maintaining the inversion symmetry. The threshold vanishes for unequal diffusion probabilities, which means that the system has a consensus state for all values of population densities. The stationary collective opinion is dominated by the individuals that diffuse more.

\end{abstract}


\maketitle

\section{Introduction}

We consider a modified definition of a widely studied consensus formation model, namely the Majority Vote model\cite{Oliveira-1992, Pereira-2005, Yu-2017, Wu-2009, Crochik-2005, Vilela-2009, Lima-2012, Vieira-2016, Fronczak-2017, Krawiecki-2018, Stanley-2018, Alves-2019}. Here, we want to study the effects of itinerant spins combined with a local majority rule\cite{Galam-2008-2} in the model dynamics. We call this definition as Diffusive Majority Vote (DMV) model, and it is a reaction-diffusion process\cite{vanKampen-1981, Dickman-1999, Odor-2004}. In particular, we are interested in the effect of different diffusive taxes for itinerant spins on the consensus formation as a way to break the inversion ($\mathbb{Z}_2$) symmetry.

We introduced a population of individuals, each one attached with a spin with values $\sigma=\pm 1$, that are allowed to hop between neighboring nodes with different probabilities $D_+$, and $D_-$. The control variable is the density $\rho$, defined as the mean number of individuals per node. In the symmetric situation, where $D_+ = D_-$ we have the inversion symmetry, and we expect a ferro-paramagnetic transition for a local update rule. However, it is not clear at first sight what it happens when we broke the $\mathbb{Z}_2$ symmetry by turning the probabilities $D_+ \ne D_-$. Symmetry breaking, in this case, means favoring an opinion value, which can describe individuals more avid to spread its opinion.

Another question is if the Ising universality class is robust when combining the local update rule with diffusion. It is known that the models with local updates given by a contact process, when transport is dominated by brownian diffusion, can have different critical exponents in lower dimensions. The Diffusive Epidemic Process (DEP)\cite{Wijland-1998, Jansen-2001, Fulco-2001, Maia-2007, Costa-2007, Dickman-2008, Argolo-2009, daSilva-2013, Tarpin-2017, Alves-2021} is an example of a system that presents a new universality class, where exponents in lower dimensions deviate from the exponents of the contact process (CP). The CP obeys the directed percolation (DP) universality class, and DEP in lower dimensions defines new universality classes\cite{Tarpin-2017}. The DEP, and DP universality classes have the same set of critical exponents in the Mean-Field regime.

In summary, our main objective is to study the critical behavior of the DMV model when changing the diffusive probabilities, and noise parameter value, which enters in the local update rule. In particular, the system presents a consensus phase for densities greater than a threshold value, which depends directly on the noise parameter. In the section 2, we present the model definition, and the relevant observables. In section 3, we discuss the main results. Finally, in section 4, we present our conclusions.

\section{Model, and Scaling}

\subsection*{The DMV Model}

In the following, we present our definition of the DMV model. We consider a population $W$ of walkers, given in terms of the concentration $\rho$ as
\begin{equation}
  W = \rho N
\end{equation}
where $N$ is the number of the lattice nodes. We assign a spin variable to each walker, which can assume two values $\sigma=\pm1$. The markovian chain is defined by the following rules:
\begin{enumerate}
\item \textbf{Initialization:} Population is randomly distributed in the lattice, and every spin can randomly assume two values $\sigma=\pm 1$, and diffuse with respective probabilities $D_\pm$. We store the number of spins $+1$, and $-1$ in each node by using two arrays $S_\pm$ of size $N$. Every dynamics step is done in a unitary time interval, and are composed of two stages:
\item \textbf{Diffusion:} All nodes are visited, and for each individual with spin $\sigma=\pm 1$ in the node $i$, one should generate a random uniform number $x$, and if $x\leq D_\pm$, the individual jumps to a randomly chosen neighbor $j$, in such a way that the arrays are updated as follows
     \begin{eqnarray}
        S_\pm(i) &=& S_\pm(i) - 1, \nonumber \\
        S_\pm(j) &=& S_\pm(j) + 1.
     \end{eqnarray}
\item \textbf{Reaction:} Inside a node, the spins follow the two-state MV model update rule\cite{Oliveira-1992, Pereira-2005, Yu-2017}. All nodes are visited, and for each spin $\sigma=\pm 1$ in the node $i$, we try a spin flip with a probability $\omega_\pm$, written as
\begin{equation}
   \omega_\pm = \frac{1}{2} \left[1 \pm (1-2q)\Theta\left( S_+(i) - S_-(i) \right)\right],
\label{mvfliprate}
\end{equation}   
where $\Theta(x)$ is the \textit{signal} function, associated with the node majority opinion
\begin{equation}
\Theta(x) = \left\lbrace \begin{array}{cl}
-1, & \mathrm{if\ }x<0; \\
0,  & \mathrm{if\ }x=0; \\
1,  & \mathrm{if\ }x>0.
\end{array}
\right.
\end{equation}
Note that the definition can consider only the other spins in the same node (excluding the spin we are trying the spin flip) when computing the local majority, however, this has no effect on the critical behavior of the model. In the case of no local majority ($x=0$ or $S_+(i)=S_-(i)$), the spin can change its opinion state with $\omega_\pm=1/2$. Noise parameter induces a no-consensus phase, analogous to the paramagnetic phase of magnetic materials. We repeat rules 2, and 3 by a number of predefined Monte Carlo (MC) steps, and for every repetition, we can increase a time counter by one unit.
\end{enumerate}

\subsection*{Observables, and critical behavior}

After describing the DMV dynamics, we present the needed observables to identify their critical behavior. The main observable is the magnetization
\begin{equation}
m = \frac{1}{W} \sum_i^N \left[S_+(i) - S_-(i)\right] .
\end{equation}
From the moments of the time series of $m$, we can obtain the order parameter $M$, its respective susceptibility $\chi$, and Binder's fourth-order cumulant $U$, which are given by\cite{Oliveira-1992}
\begin{eqnarray}
M(q)&=& \langle \left\vert m \right\vert \rangle, \nonumber \\
\chi(q) &=& N \left( \langle m^{2} \rangle - \langle \left\vert m \right\vert \rangle ^{2} \right), \nonumber \\
U(q) &=& 1 - \frac{\langle m^{4} \rangle}{3\langle m^{2} \rangle^{2}},
\label{observables}
\end{eqnarray}
respectively, where $\left\vert m \right\vert$ is the absolute value of $m$. All observables are functions of noise parameter $q$, and concentration $\rho$. 

The observables written in Eq.(\ref{observables}) should obey the following finite-size scaling (FSS) relations
\begin{eqnarray}
M &=& L^{-\beta/\nu} f_{M}\left( L^{1/\nu} \left(\rho - \rho_{c}\right) \right), \nonumber \\
\chi &=& L^{\gamma/\nu} f_{\chi}\left( L^{1/\nu} \left(\rho - \rho_{c}\right) \right), \nonumber \\
U &=& f_{U}\left( L^{1/\nu} \left(\rho - \rho_{c}\right) \right),
\label{observables-fss}
\end{eqnarray}
where $L$ is the linear size of the lattice. The number of nodes is $N=L^d$, where $d$ is the dimension of the lattice. In the scaling relations given by Eq.(\ref{observables-fss}), $1/\nu$, $\beta/\nu$, and $\gamma/\nu$ are the critical exponent ratios, $\rho_c$ is the critical noise, and $f_{M,\chi,U}$ are the finite-size scaling functions.

To obtain the relevant observables, we performed the dynamics on square lattices with sizes $L=50$, $L=60$, $L=70$, $L=80$, $L=90$, and $L=100$, and cubic lattices with sizes $L=14$, $L=16$, $L=18$, $L=20$, $L=22$, and $L=24$. In all simulations shown here, we used a noise value of $q=0.1$. In general, for a finite value of noise parameter, we obtain a threshold for $D_+=D_-$ that increases with the noise parameter. We considered $10^6$ MC steps in order to evolve the system in a stationary state, and another $10^7$ MC steps to collect $10^7$ values of the opinion balance to measure the observables. Statistical errors were calculated by using the ``jackknife'' resampling technique\cite{Tukey-1958}. 

\section{Results, and Discussion}

\begin{figure}[p]
\begin{center}
\includegraphics[scale=0.17]{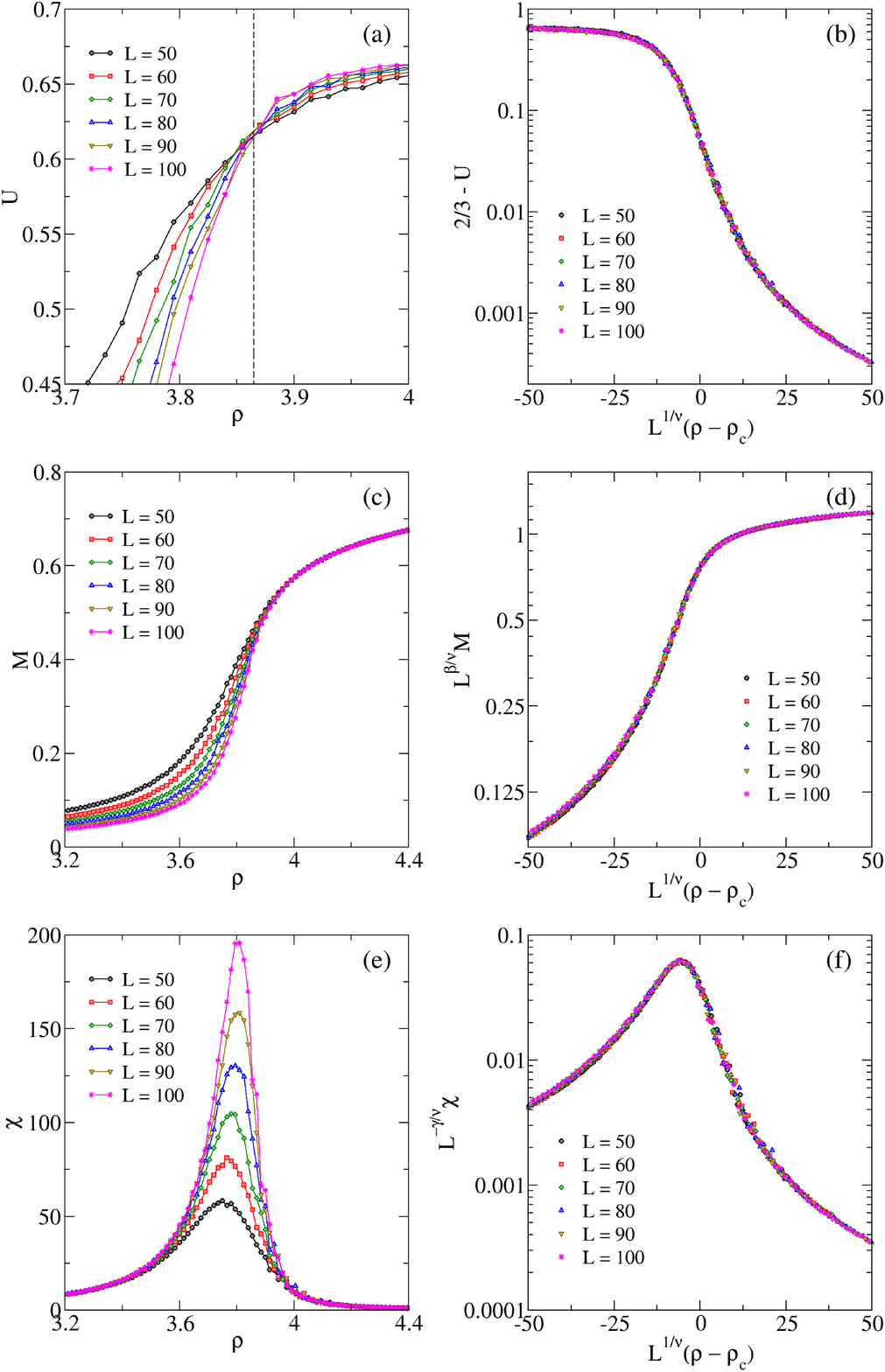} 
\end{center}
\caption{(Color Online) Results of the averages (\ref{observables}) for the DMV model on a square lattice, with diffusion probabilities $D_+=D_-=0.5$, and noise parameter $q=0.1$. In panels (a), (c), and (e), we show our numerical data for Binder cumulant $U$, the order parameter $M$, and susceptibility $\chi$ for   different lattice sizes. In panels (b), (d), and (f) we show the respective data colapses following Eq.(\ref{observables-fss}), and the critical exponents given in Tab.(\ref{criticalbehaviortable}) for 2d. The estimated value of the critical threshold for the square lattice is $\rho_c=3.862(5)$. Statistical errors are smaller than symbols.}
\label{results-d=2-D+=D-=0.5}
\end{figure}

\begin{figure}[p]
\begin{center}
\includegraphics[scale=0.17]{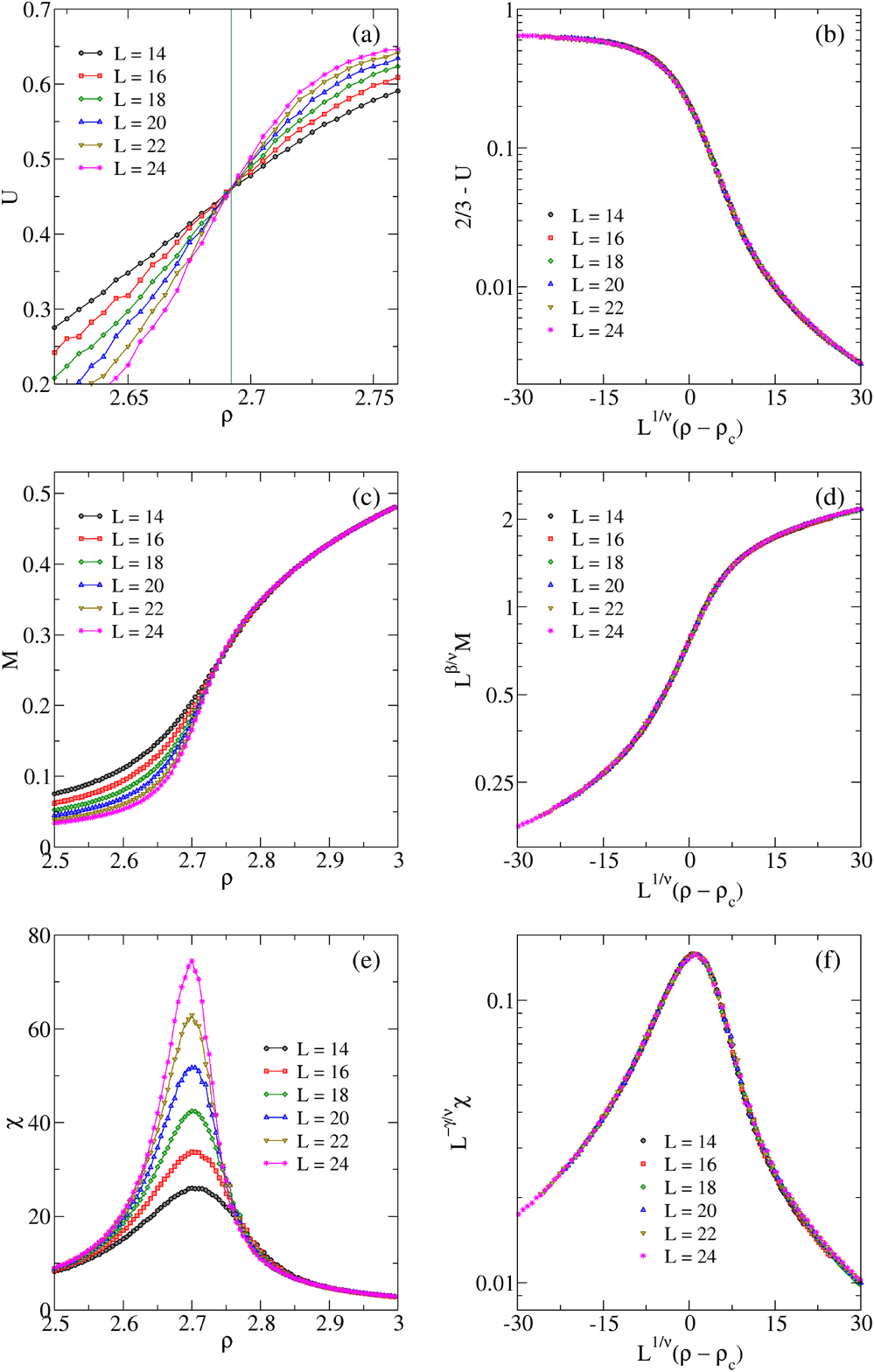} 
\end{center}
\caption{(Color Online) The same as Fig.(\ref{results-d=2-D+=D-=0.5}) for the simple cubic lattice. The estimated value of the critical threshold for the simple cubic lattice is $\rho_c=2.692(5)$. The data colapses follow Eq.(\ref{observables-fss}), and the critical exponents given in Tab.(\ref{criticalbehaviortable}) for 3d. Statistical errors are smaller than symbols.}
\label{results-d=3-D+=D-=0.5}
\end{figure}

We start by present the results for the relevant observables of the DMV model. In Fig.(\ref{results-d=2-D+=D-=0.5}), and (\ref{results-d=3-D+=D-=0.5}) we show the results for the cumulant, the order parameter, and its susceptibility for square, and cubic lattices, respectively, for q=$0.1$, and $D_+=D_-=0.5$. For a finite value of noise parameter, and equal diffusion probabilities $D_+=D_-$, the system undergoes a continuous phase transition by increasing concentration.

The threshold increases with the value of noise parameter, indicating that an increasing noise strength induces the paramagnetic phase by disorder, and on the other hand, the increasing density induces consensus (paramagnetic order). In addition, the threshold depends on the value of diffusive probabilities $D_+=D_-$ too. When increasing the value of the diffusive probabilities, the threshold decreases.

By the dependence of the density threshold on the diffusive probabilities, we can conclude that the two main mechanisms to induce consensus in this model are concentration, and diffusion, as expected. The model is compatible with the fact that isolation leads to dissensus in a population of individuals. 

Regarding the scaling behavior, the continuous transition is in the Ising universality class. This is unexpected at first glance because, as already mentioned in the introduction, the reaction-diffusion version of the contact process, known as the DEP, have a different set of exponents from the CP defined on a sedentary population. According to the Janssen-Grassberger conjecture\cite{Janssen-1981, Grassberger-1982}, the particle conservation laws of the DEP are responsible for changing universality class. However, the introduction of diffusion while maintaining $\mathbb{Z}_2$ symmetry still leave the universality class of a system in the Ising universality class.

In two dimensions (2d), the expected value of $\nu$ for the DEP with brownian diffusion is $2/d=1$\cite{Kree-1989, Jansen-2001}, which is the same as Ising model in 2d, and this could lead, in principle, to unchanged values for all exponents in 2d with respect to the Ising universality class. To determine if the diffusion can change the universality class, we simulated the dynamics in the cubic lattice where the $\nu$ exponent for the Ising model is different from the diffusion exponent $2/d$\cite{Kree-1989, Jansen-2001}. Results shown on Fig.(\ref{results-d=3-D+=D-=0.5}) are compatible with the Ising universality class in three dimensions (3d).

\begin{table}[h]
\begin{center}
\begin{tabularx}{0.8\textwidth}{YYY}
\hline
Critical exponents & Values in 2d & Values in 3d \\
\hline
$\nu$      & $1$   & $0.629971(4)$ \\
$\beta$    & $1/8$ & $0.326419(3)$ \\
$\gamma$   & $7/4$ & $1.237075(10)$ \\
\hline                 
\end{tabularx}
\end{center}
\caption{Ising critical exponents. We used, in our data collapses, the exact Ising exponents in 2d, and the best estimations in 3d to date, given in Ref.\cite{Kos-2016}.}
\label{criticalbehaviortable}
\end{table}

\begin{figure}[h]
\begin{center}
\includegraphics[scale=0.24]{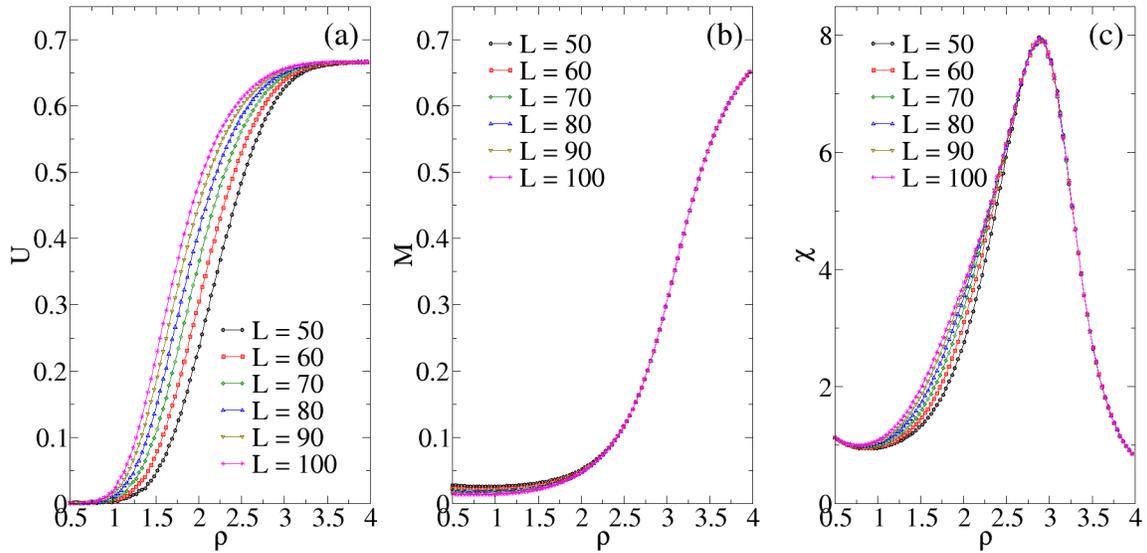} 
\end{center}
\caption{(Color Online) The same as Fig.(\ref{results-d=2-D+=D-=0.5}) except $D_-=0.55$. The cumulant does not present any crosses, indicating that the DMV with different diffusion probabilities does not have a critical threshold. The magnetization increases with the density, and is not zero, which is compatible with a net magnetization. The susceptibility does not diverge in the infinite lattice limit, instead, it presents a finite maximum at $\rho=2.90(5)$.}
\label{results-d=2-D+=D-=0.55}
\end{figure}

In the following, we analyze what happens if one chooses different values for the diffusion probabilities for each spin state in order to break $\mathbb{Z}_2$ symmetry. In Fig.(\ref{results-d=2-D+=D-=0.55}) we show the cumulant, and magnetization for a DMV evolution with $D_+=0.5$, and $D_-=0.55$. We note that the cumulant does not present any crossings in a way that the critical threshold is destroyed. From the magnetization, we see that the system presents a net magnetization for all densities, and the magnetization increases with concentration.

\begin{figure}[h]
\begin{center}
\includegraphics[scale=0.22]{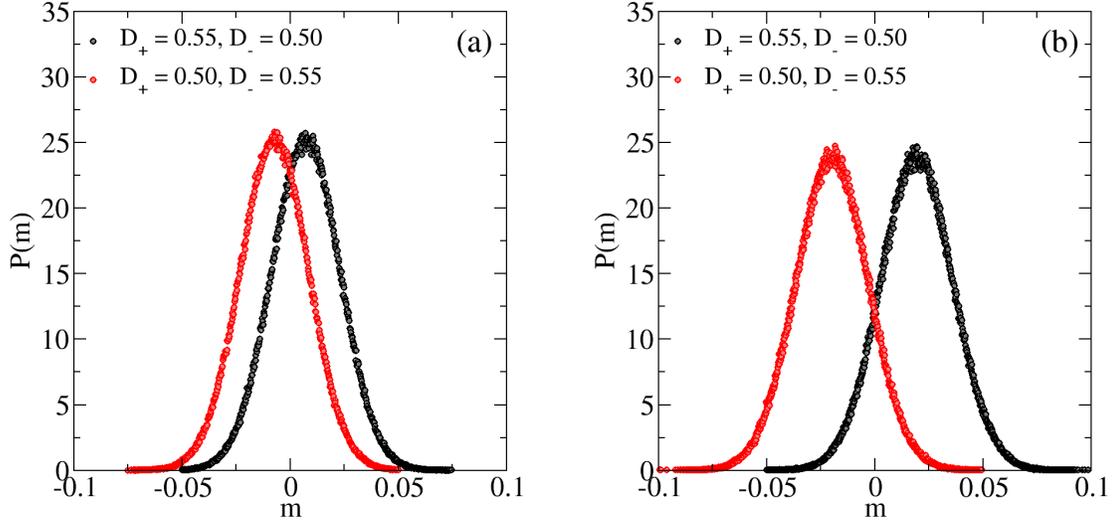} 
\end{center}
\caption{(Color Online) In panel (a) we show two histograms of the time series of m for the stationary evolution of DMV on a square lattice with $D_+=0.55$, and $D_-=0.5$ (black circles), and with $D_+=0.5$, and $D_-=0.55$ (red circles). For both data sets, noise parameter is $q=0.1$, $L=100$, and concentration is $\rho=1$. In panel (b) we have the same as the panel (a) except the concentration, which is $\rho=1.5$. Note that the sign of the most probable value of $m$ follows the one of the most diffusing particles. The most probable value of $m$ increases with the density which can be seen from the spacing between the maxima of the histograms in panels (a), and (b).}
\label{histogram-rho=1.5}
\end{figure}

The system becomes ferromagnetic for all finite densities, as we can see from the histogram of the time series with $10^6$ values of $m$ for a DMV process with different diffusive probabilities, shown in Fig.(\ref{histogram-rho=1.5}), where the most probable value is not $m=0$ anymore, and instead, the net magnetization is dominated by the most diffusing particles. Interchange of different values reflects the histogram, changing the sign of the most probable value. In addition, the maxima separation of the histograms from $m=0$ increases with concentration in a way that if we increase the population of walkers, we produce a greater net magnetization.

\begin{figure}[h]
\begin{center}
\includegraphics[scale=0.10]{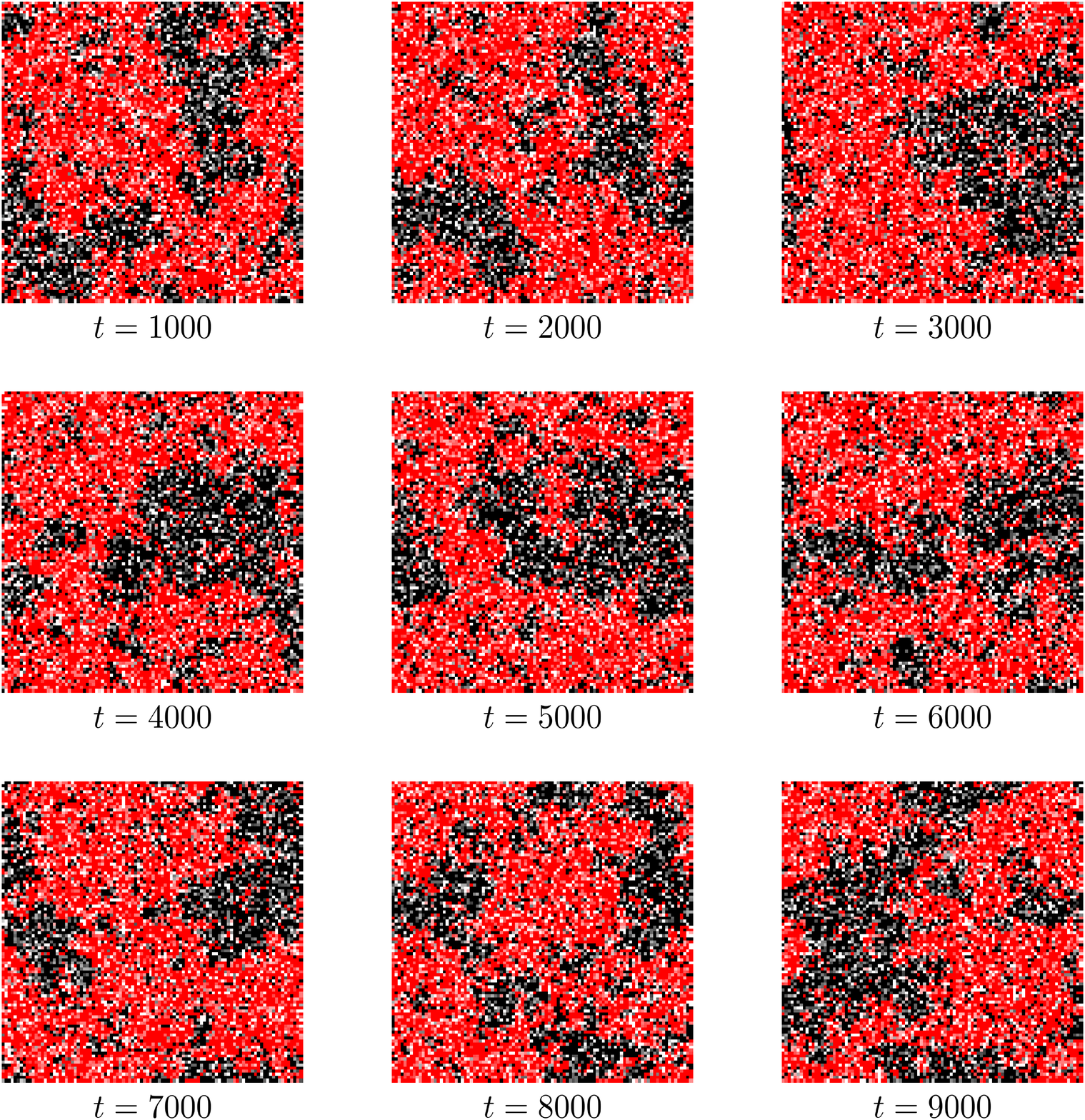}
\end{center}
\caption{(Color Online) Snapshots of DMV dynamics on a square lattice, with diffusion probabilities $D_+=D_-=0.5$, and noise parameter $q=0.1$. A node is black if the majority of its spins have the value $1$, is red for the opposite, and is white for no local majority. We simulated the dynamics for the value of the susceptibility maxima $\rho=3.81(5)$, and note the presence of clusters, where walkers with the same polarization tend to stay together at the transition to the ferromagnetic phase.}
\label{snapshots-d+=0.5-d-=0.5}
\end{figure}

Now, we show snapshots of the DMV dynamics. First, in Fig.(\ref{snapshots-d+=0.5-d-=0.5}), we present the dynamics for equal diffusive probabilities $D_+=D_-=0.5$ on the maxima of fluctuations for a square lattice, with $L=100$, estimated at $\rho=3.81(5)$. Note that the dynamics generate clusters of populations with a defined polarization state, where spins with the same polarization tend to stay together. When close to the critical threshold, the system has clusters of various sizes.

\begin{figure}[h]
\begin{center}
\includegraphics[scale=0.10]{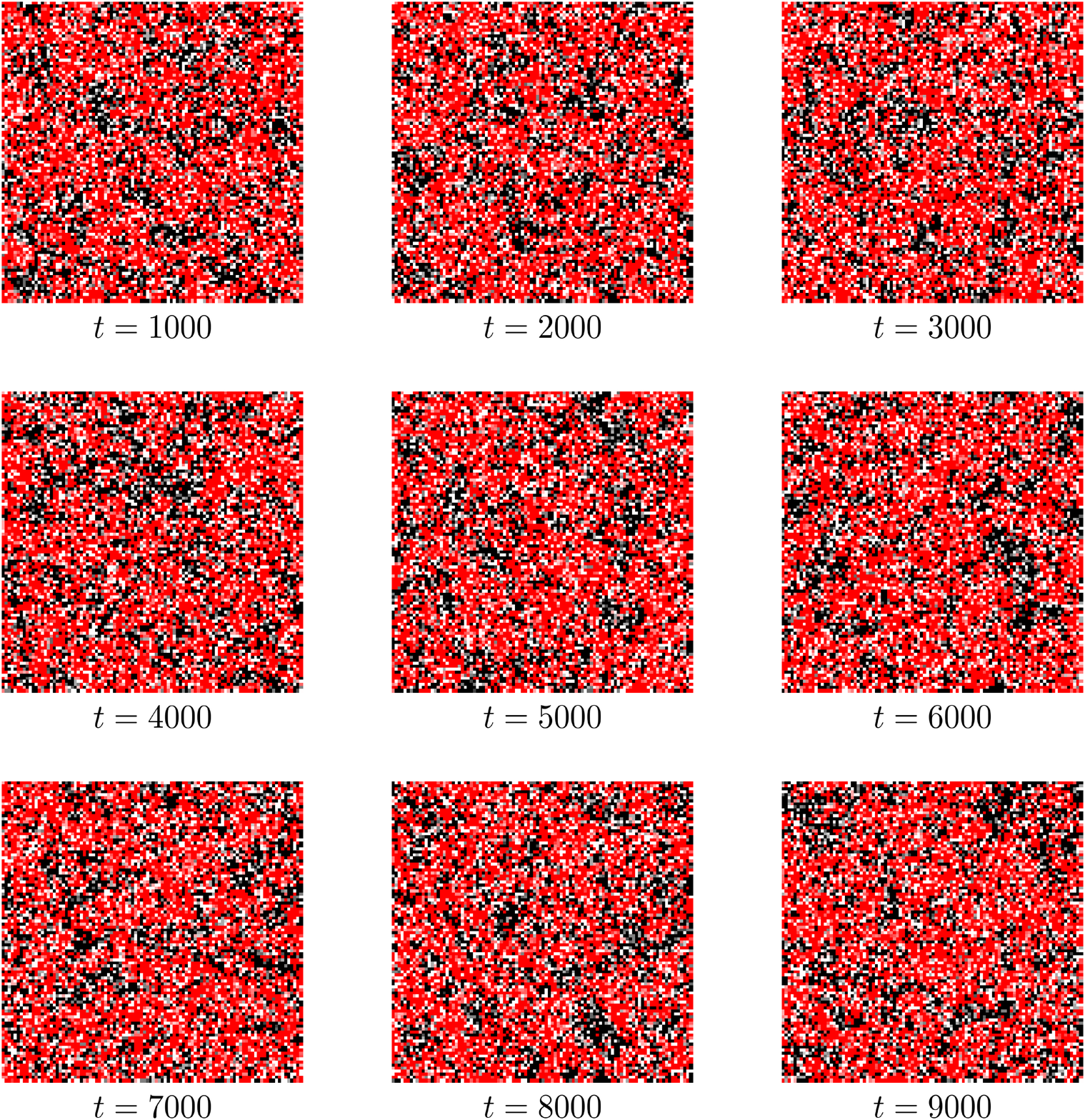} 
\end{center}
\caption{(Color Online) Snapshots of DMV dynamics with the same parameters of Fig.(\ref{snapshots-d+=0.5-d-=0.5}) except $D_-=0.55$, and the value of the susceptibility maxima that is $\rho=2.90(5)$. Note that the walkers with the same polarization are now more dispersed, and that the more diffusing population is dominant.}
\label{snapshots-d+=0.5-d-=0.55}
\end{figure}

In Fig.(\ref{snapshots-d+=0.5-d-=0.55}), we show the same for $D_+=D_-=0.55$ on the maxima of fluctuations, now on $\rho=2.90(5)$. Differently of the equal diffusion probabilities case, we see populations of spins $+1$, and $-1$ more dispersed. In addition, the individuals with spins $\sigma=-1$ are dominant, which are the most diffusing ones.

\section{Conclusions}

We presented a consensus formation model in a system composed of non-sedentary individuals that can interact in the same node of a lattice/network according to the majority rule. The model leaves the possibility of assign different diffusion probabilities for the different individual opinion states, and the presence of a random noise that describes Galam contrarians\cite{Galam-2008-1}. The DMV is a reaction-diffusion process, where the transport is dominated by brownian diffusion, and the reactions are done by following a local majority rule.

In the case of $\mathbb{Z}_2$ symmetry, i.e., equal diffusion probabilities, the system presents a continuous phase transition from a paramagnetic phase to a ferromagnetic state, with global consensus when increasing the number of individuals. In this way, the model predicts that isolation can potentialize local disagreements, and induce a global dissensus state. Increasing concentration induces a contrary effect of the Galam contrarians, i.e., preserving the global consensus. As expected, the critical threshold increases with the noise parameter, and decreases with the equal diffusion probabilities.

When breaking the $\mathbb{Z}_2$ symmetry, we destroy the critical threshold, and the system presents a global consensus state in the infinite lattice limit for any population. The dominant (consensus) opinion state is determined by the greater diffusion probability. This is consistent with the fact that the dominant opinion state is the opinion state of the individuals that are more eager to convince the others.

\section{Acknowledgments}

We would like to thank CAPES (Coordena\c{c}\~{a}o de Aperfei\c{c}oamento de Pessoal de N\'{\i}vel Superior), CNPq (Conselho Nacional de Desenvolvimento Cient\'{\i}fico e tecnol\'{o}gico), FUNCAP (Funda\c{c}\~{a}o Cearense de Apoio ao Desenvolvimento Cient\'{\i}fico e Tecnol\'{o}gico), and FAPEPI (Funda\c{c}\~{a}o de Amparo a Pesquisa do Estado do Piau\'{\i}) for the financial support. We acknowledge the use of Dietrich Stauffer Computational Physics Lab, Teresina, Brazil, and Laborat\'{o}rio de F\'{\i}sica Te\'{o}rica e Modelagem Computacional - LFTMC, Teresina, Brazil, where the numerical simulations were performed.

\bibliography{textv1}

\end{document}